\documentclass[aps,showpacs,twocolumn,prl]{revtex4}
\usepackage{graphicx}
\usepackage{amsmath}
\begin{document}
\title{Quantum Faraday Effect in Double-Dot Aharonov-Bohm Ring}
\author{Kicheon Kang}
\affiliation{Department of Physics, Chonnam National University,
 Gwangju 500-757, Republic of Korea,}
\affiliation{Institute of Experimental and Applied Physics, University
 of Regensburg, 93040 Regensburg, Germany}
\date{\today}
\begin{abstract}

We investigate Faraday's law of induction manifested in the
quantum state of Aharonov-Bohm loops. In particular, we propose
a flux-switching experiment for a double-dot AB ring
to verify the phase shift induced by Faraday's law. We show that
the induced {\em Faraday phase} is geometric and nontopological.
Our study demonstrates that the relation between the local
phases of a ring at different fluxes is not arbitrary but is instead
determined by Faraday's inductive law, which is in strong contrast to the
arbitrary local phase of an Aharonov-Bohm ring for a given flux.

\end{abstract}
\pacs{73.23.-b, 
      03.65.Vf, 
      03.65.Wj 
     }
\maketitle

We begin by pointing out an apparent paradox between
the two well known facts in quantum theory:
(1) the local phase factor of the wave function
in an Aharonov-Bohm (AB) loop is arbitrary~\cite{gasiorowicz};
(2) the wave function (more generally, the density matrix) of a quantum system
can be reconstructed by a technique of the quantum state tomography (QST)
on repeated preparation of the system~\cite{vogel89,paris04}.
Aharonov-Bohm (AB) effect~\cite{aharonov59} is one of the most striking
phenomena discovered in quantum theory.
In the case of an arbitrary AB loop at equilibrium,
any physical quantity is periodic in $\Phi$ with a period
$\Phi_0=e/hc$, the flux quantum (``Byers-Yang's theorem")~\cite{byers61}.
%
AB effect is gauge-invariant and appears
as a manifestation of the gauge-invariant phase factor.
The choice of a particular gauge is arbitrary, and therefore the local phase
factor of the wave function is also arbitrary.

It is interesting to note that this arbitrary local phase factor of an AB loop is
inconsistent with the
fact that the wave function can be reconstructed by the technique of QST.
The gap between the two facts needs to be clarified.
In this Letter, we show that the law of Faraday's induction plays
a central role when we try to reconstruct the wave function of an AB loop.
For a full reconstruction of the wave function, change of the AB flux is
inevitable. The change of the AB flux results in Faraday's induction,
and gives rise to an additional phase shift. We find that this Faraday-induced
phase shift is geometric and nontopological. It is geometric in the sense
that it depends only on the net change of the magnetic field. On the other
hand, it is
nontopological since it depends not only on the flux change but also
on the specific geometry of the system.

We analyze the characteristics of the {\em quantum Faraday effect} for a
tunnel-coupled double-dot AB loop with a localized magnetic flux
$\Phi$ penetrating the hole (Fig.~1(a)). This is the simplest two-state
problem involving an AB flux.
The Hamiltonian of a double-dot loop is given by
\begin{subequations}
\label{eq:hamil-dd}
\begin{equation}
 H = \sum_{\alpha=1,2}\varepsilon_\alpha c_\alpha^\dagger c_\alpha
   - (t e^{i\phi_a} c_1^\dagger c_2 + t e^{i\phi_b} c_2^\dagger c_1)
   - \mbox{h.c.} ,
\end{equation}
where $c_\alpha$($c_\alpha^\dagger$) annihilates(creates) an electron
at QD-$\alpha$ ($\alpha=1,2$) with its energy level
$\varepsilon_\alpha$.
For simplicity, tunneling amplitude $t$ is
assumed to be identical for both the upper (path $a$) and the lower
(path $b$) paths.
The AB phase $\phi$ ($=2\pi\Phi/\Phi_0$) is given by $\phi=\phi_a+\phi_b$.
The ``local" phases $\phi_a$ and $\phi_b$ are not
gauge-invariant. Only $\phi$ is a gauge-invariant phase.
This Hamiltonian can be simplified as
\begin{equation}
 H = \sum_{\alpha=1,2}\varepsilon_\alpha c_\alpha^\dagger c_\alpha
   + t_\phi c_1^\dagger c_2 + t^*_\phi c_2^\dagger c_1 ,
\end{equation}
with the effective tunneling amplitude
\begin{equation}
 t_\phi = -2te^{i(\phi_a-\phi_b)}\cos{(\phi/2)}\,.
\label{eq:tphi}
\end{equation}
\end{subequations}
Note that $t_\phi$ depends on
the choice of gauge due to the phase factor $e^{i(\phi_a-\phi_b)}$.

It is useful to rewrite the Hamiltonian (Eq.~\ref{eq:hamil-dd}) in
a Bloch-sphere (pseudospin) representation with $|\uparrow\rangle=|1\rangle$ and
$|\downarrow\rangle=|2\rangle$:
\begin{subequations}
\begin{equation}
 H = -\frac{1}{2} \vec{\sigma}\cdot\mathbf{B} ,
\end{equation}
where
\begin{equation}
 \mathbf{B} = (-2\mbox{\rm Re}(t_\phi),2\mbox{\rm Im}(t_\phi),
   \Delta\varepsilon)
\label{eq:B}
\end{equation}
\end{subequations}
is a pseudo-magnetic field,
and $\vec{\sigma} = (\sigma_x,\sigma_y,\sigma_z)$.
$\Delta\varepsilon=\varepsilon_2-\varepsilon_1$ is the energy
level detuning of the two dots. We imposed the condition
$\varepsilon_1+\varepsilon_2=0$ without loss of generality.

It is straightforward to obtain the eigenstate energies
and the corresponding eigenvectors.
Although a multi-valued wave function cannot be ruled
out {\em a priori}~\cite{byers61}, we do not consider this possibility
because a QST with multi-valued wave function is meaningless~\cite{merzbacher62}.
The two eigenstate energies are
$E_\pm(\mathbf{B})=\mp \frac{1}{2} |\mathbf{B}|
  = \mp\frac{1}{2} \sqrt{(\Delta\varepsilon)^2+4|t_\phi|^2}$, with the
corresponding eigenvectors being given by
\begin{subequations}
\label{eq:psi-dd}
\begin{equation}
 |\pm(\mathbf{B})\rangle = \alpha_\pm|\uparrow\rangle
    + \beta_\pm|\downarrow\rangle ,
\end{equation}
where
\begin{equation}
 \alpha_\pm =
  \frac{t_\phi}{ \sqrt{ (\varepsilon_1-E_\pm)^2+|t_\phi|^2} },
  \; \beta_\pm =
  -\frac{\varepsilon_1-E_\pm}{ \sqrt{ (\varepsilon_1-E_\pm)^2 +|t_\phi|^2} }.
\label{eq:alpha}
\end{equation}
\end{subequations}
Although the energy eigenvalue $E_\pm$ is periodic in the AB phase $\phi$
with a period of $2\pi$ (as indicated by Byers-Yang's
theorem~\cite{byers61}),
the wave function does not show such periodicity.
Instead, the wave function depends on the arbitrary choice of the
phases $\phi_a$ and $\phi_b$.
This is evident from the gauge-dependent factor $t_\phi$ in
Eq.~(\ref{eq:alpha}) (see also Eq.~(\ref{eq:tphi})).


Now, let us discuss the following flux-switching and pseudospin precession
experiment (see also the illustration of Fig.~1(b)).
This kind of time-domain experiment is an essential part
of a QST with a system involving an AB flux~\cite{liu05}.
Initially, the AB flux is prepared to have an arbitrary, fixed
value $\phi$, and
the system is in the ground state $|+(\mathbf{B})\rangle$ of Eq.~(\ref{eq:psi-dd}).
It is assumed that thermal fluctuations are small enough that
the system is in the ground state, that is, $k_BT\ll|\mathbf{B}|$.
Then, the external magnetic flux is suddenly dropped to zero,
and the pseudo-magnetic
field $\mathbf{B}$ changes immediately to the value given by
Eq.~(\ref{eq:B}) with $\phi=0$ and
$\phi_b=0$. ($\phi_b$ might be chosen to be nonzero in general, but it does not
make any change to our findings here.) This corresponds to
\begin{equation}
 \mathbf{B} \rightarrow \mathbf{B}_0 = (4t,0,\Delta\varepsilon) .
\end{equation}
Because of the sudden change in $\mathbf{B}$, the state of
the electron, denoted by $|\psi_{\phi\rightarrow0}(t)\rangle$,
will precess according to the relation (shown as a dashed line in Fig.~1(b))
$ |\psi_{\phi\rightarrow0}(t)\rangle
   = e^{\frac{i}{2}\vec{\sigma}\cdot\hat{n}B_0 t} |+(\mathbf{B})\rangle$,
where $\hat{n}$ is the unit vector parallel to $\mathbf{B}_0$, and $B_0$ is its magnitude. We obtain
\begin{subequations}
\begin{equation}
  |\psi_{\phi\rightarrow0}(t)\rangle
    = \alpha_0(t)|\uparrow\rangle + \beta_0(t)|\downarrow\rangle ,
\end{equation}
where
\begin{eqnarray}
 \alpha_0(t) &=& \alpha_+\cos{\frac{B_0t}{2}}
    + i\left( \alpha_+ n_z+\beta_+ n_x \right) \sin{\frac{B_0t}{2}}, \\
 \beta_0(t) &=& \beta_+\cos{\frac{B_0t}{2}}
    - i\left( \beta_+ n_z-\alpha_+ n_x \right) \sin{\frac{B_0t}{2}}.
\end{eqnarray}
\end{subequations}
Here $n_x$($n_z$) is the $x$($z$) component of the vector $\hat{n}$.

The time evolution of the state, $|\psi_{\phi\rightarrow0}(t)\rangle$,
depends on the gauge, simply because
$\mathbf{B}$ depends on the gauge.
Furthermore, this gauge dependence is shown in physical quantities, for instance,
in the time evolution of
the electron number at QD-1 (or at QD-2):
\begin{equation}
  n_1(t) =
  \langle\psi_{\phi\rightarrow0}(t)
   |c_1^\dagger c_1|\psi_{\phi\rightarrow0}(t)\rangle
  = |\alpha_0(t)|^2 .
\end{equation}
The time evolution of $n_1$ is displayed for
two different choices of gauge:
(i) symmetric gauge ($\phi_a=\phi_b=\phi/2$) (Fig.~2), and (ii)
``$\phi_b=0$" gauge (Fig.~3).
(In fact, an infinite number of gauge choices exists which satisfy the
condition $\phi_a+\phi_b=\phi$). For the symmetric gauge,
the effective
tunneling amplitude is $t_\phi=-2t\cos{(\phi/2)}$.  This gauge
leads to a $4\pi$-periodicity of the Hamiltonian
and its eigenstates (Eqs.~(\ref{eq:hamil-dd},\ref{eq:psi-dd})).
On the other hand, for the ``$\phi_b=0$" gauge,
$t_\phi=-t(e^{i\phi}+1)$, which provides a $2\pi$-periodicity of the Hamiltonian
and its eigenstates. Obviously, a different choice of gauge leads to a different
result.

Of course, the gauge dependence of the results shown in Fig.~2 and Fig.~3
should not
exist in reality. The result should be unique in spite of the choice of gauge.
In fact, the contradiction is resolved if we take into account
Faraday's law induction in the flux-switching procedure. When the localized
magnetic field changes in time, the electric field
\begin{equation}
 \mathbf{E} = -\nabla V - \frac{1}{c}\frac{\partial\mathbf{A}}{\partial t}
\end{equation}
is induced, where $V$ and $\mathbf{A}$ are the scalar and the vector potentials,
respectively. The choice of the gauges for $V$ and $\mathbf{A}$ should provide
the correct value of $\mathbf{E}$, which was not taken into account in obtaining
the results displayed Figs.~2 and 3. Still there is a freedom to choose the gauge,
and we choose $V$ to be time-independent, and then the inductive component of the
field, $\mathbf{E}_t$, is given by
$\mathbf{E}_t = - \frac{1}{c}\frac{\partial\mathbf{A}}{\partial t}$.
For the system under consideration, the following
relations are derived from Faraday's law:
\begin{equation}
 \int_\gamma \mathbf{E}_t\cdot d\mathbf{R} =
 -\frac{1}{c}\frac{\partial}{\partial t} \int_\gamma \mathbf{A}\cdot d\mathbf{r}
  = -\frac{\hbar}{e}\frac{\partial\phi_\gamma}{\partial t},
\end{equation}
where $\int_\gamma$ represents an integral over the path $\gamma$
($a$ or $b$).

The question at this point is how to choose a gauge giving the
correct inductive field. Actually, the induced electric field depends not
only on the time-dependent flux $\Phi(t)$
but also on the specific geometry of the system and the distribution of the
localized magnetic flux. Let us consider, for example,
a highly symmetric
double-dot ring: with circular symmetric AB flux and identical paths for the
upper (path $a$) and the lower (path $b$) parts of the ring. Then, the symmetry of
the system leads to the relation
$\int_a\mathbf{E}_t\cdot d\mathbf{R}=\int_b\mathbf{E}_t\cdot d\mathbf{R}$,
or
 $\frac{\partial\phi_a}{\partial t} = \frac{\partial\phi_b}{\partial t}$.
We find that this condition is fulfilled by choosing the symmetric gauge
for the time-dependent phases: $\phi_a(t)=\phi_b(t)=\phi(t)/2$.
Therefore, the results shown in Fig.~2 are correct for the symmetric system
because they give
the correct value of $\mathbf{E}_t$.
In general, the gauge should be selected to give the correct value of
$\mathbf{E}_t$, which depends on the specific geometry of the system.

An important implication of the above discussion on Faraday's induction
is that it gives an additional local phase in the wave function of the system.
In the following, we show that Faraday's induction gives a
geometrical
(but nontopological) phase shift. Here we discuss it for a specific
double-dot ring
system, but it can be applied equally to any AB loops. The initial (local) phases
at $t=t_i$ are represented by
$\phi_a^i$ and $\phi_b^i$ for paths $a$ and $b$, respectively.
These phases evolve as the magnetic field changes in time, and the final
values (at $t=t_f$) are given by $\phi_a^i+\Delta\phi_a$
and $\phi_b^i+\Delta\phi_b$,
respectively.  During the change in the magnetic field,
the inductive field induces a momentum kick $\Delta \mathbf{p}(\mathbf{r})$
which depends on the position $\mathbf{r}$ as
\begin{equation}
 \Delta\mathbf{p}(\mathbf{r}) = e\int_{t_i}^{t_f}\mathbf{E}_t\, dt
  = -\frac{e}{c}\Delta\mathbf{A}(\mathbf{r}) ,
\end{equation}
where $\Delta\mathbf{A}(\mathbf{r})$ is the change in the vector potential.
This momentum kick
induces a local phase shift
\begin{equation}
 \phi^{F}(\mathbf{r}) = \frac{1}{\hbar}\Delta\mathbf{p}(\mathbf{r})\cdot
 \mathbf{r} ,
\end{equation}
in the wave function.
From this relation, one can find that Faraday's induction gives the relative
phases of the two quantum dots
\begin{displaymath}
 \phi^F_a = \frac{1}{\hbar} \int_a
  \Delta\mathbf{p}(\mathbf{r})\cdot d\mathbf{r}
  = -\Delta\phi_a ,
\end{displaymath}
for path $a$. Similarly, it gives $\phi^F_b = -\Delta\phi_b$ for path $b$.
It is interesting to note that the {\em Faraday phase} for one loop is equivalent
to the negative of the change in the AB phase:
$\phi^F_{one\,loop} = \frac{1}{\hbar} \oint
  \Delta\mathbf{p}(\mathbf{r})\cdot d\mathbf{r} = -\Delta\phi$ .
In contrast to the AB effect, not only the phase for one loop but also
the local phase $\phi^{F}(\mathbf{r})$ is
physically meaningful because the latter is directly related to
the inductive field, a physical quantity.
Note that, the local phase $\phi^{F}(\mathbf{r})$
is geometric in the sense that it depends only on the net change of
the vector potential $\Delta\mathbf{A}(\mathbf{r})$.
However, this phase is nontopological
because it depends not only on the topology of the ring, but also
on the specific geometry. The $4\pi$-periodicity of a symmetric double-dot
ring can be understood from the $4\pi$-periodicity of the phases $\phi^F_a$
and $\phi^F_b$, since it satisfies
\begin{equation}
 \phi^F_a=\phi^F_b=-\Delta\phi/2 .
\label{eq:Fab}
\end{equation}

The effect of the {\em Faraday phase} can be observed even in an adiabatic
change of the magnetic flux. Two conditions are necessary
for this purpose.
First, a nonstationary initial state should be prepared. Otherwise, the
adiabatic evolution of the magnetic field gives just an adiabatic evolution
of the ground state, and the {\em Faraday phase} is not observable.
Second, the characteristic time scale of the
flux change, denoted by $\Delta t$, should meet the condition
$\hbar/|\mathbf{B}|\ll \Delta t \ll t_{deph}$, where $t_{deph}$ is
the dephasing time of the initial nonstationary state.
The flux should change at a much slower rate than the precession frequency of
the pseudospin (adiabaticity), but
should be switched faster than the dephasing time, in order to observe
the evolution of the nonstationary state.

The procedure for a possible experiment is as follows:
(i) a ground state is prepared
with $\Delta\varepsilon=0$ and $\mathbf{B}=\mathbf{B}_i$.
(ii) A nonstationary state is initialized
by a sudden switching of the level detuning from zero
to a finite value $\Delta\varepsilon$. (iii) The magnetic flux is adiabatically
switched so that the pseudo-magnetic field $\mathbf{B}$ changes accordingly.
(iv) Finally, the time evolution
of the electron number is measured in one of the QDs.

The adiabatic process of (iii) is described by the Hamiltonian
$H\{\mathbf{B}(t)\}=-\vec{\sigma}\cdot\mathbf{B}(t)/2$
with an adiabatic change of $\mathbf{B}$.
%
The nonstationary initial state immediately after process (ii)
(for a symmetric ring with the initial AB phase in the range of
 $\pi < \phi < 3\pi$),
\begin{subequations}
\begin{equation}
 |\psi_0\rangle = \frac{1}{\sqrt{2}}
  \left( |\uparrow\rangle-|\downarrow\rangle \right) ,
\label{eq:psi0}
\end{equation}
evolves upon the adiabatic change in $\mathbf{B}(t)$. It satisfies
the adiabatic evolution of the eigenstates
\begin{equation}
 |\psi_\pm(\mathbf{B}(t)) \simeq e^{i\gamma_\pm(\mathbf{B})}
  e^{-i\int_0^{t}E_\pm(t')dt'} |\pm(\mathbf{B}(t)\rangle ,
\end{equation}
where $\gamma_\pm$ denotes the geometric phase acquired during the adiabatic
change of $\mathbf{B}(t)$~\cite{berry84}.
We find that $\gamma_\pm=0$ for the symmetric ring
(imposed by the relation $\phi_a(t)=\phi_b(t)$).
Therefore, the time evolution of the state upon adiabatic
change is given by
\begin{equation}
 |\psi(t)\rangle = c_+ e^{i\phi^+_{dy}(t)}|+(\mathbf{B}(t))\rangle
 + c_- e^{i\phi^-_{dy}(t)}|-(\mathbf{B}(t))\rangle ,
\end{equation}
where $c_\pm=\langle\pm(\mathbf{B}_i)|\psi_0\rangle$, and
\begin{equation}
 \phi^\pm_{dy}(t) \equiv -\int^t E_\pm(\mathbf{B}(t')) dt'
  = \pm \frac{1}{2}\int^t \mathbf{B}(t') dt'
\end{equation}
is a dynamical phase.
\label{eq:psi-adiabatic}
\end{subequations}
It is also useful to define the difference between the two
dynamical phases, $\Delta\phi_{dy}(t)\equiv \phi^+_{dy}(t) - \phi^-_{dy}(t)$.

The time evolution of the electron number at QD-1 (state $|\uparrow\rangle$) is
\begin{equation}
 n_1 (t) = |\langle\uparrow|\psi(t)\rangle|^2 \,.
\end{equation}
We find from Eq.~(\ref{eq:psi-adiabatic}) that
\begin{subequations}
\begin{equation}
 \langle\uparrow|\psi(t)\rangle = u(t)\cos{(\Delta\phi_{dy}(t)/2)}
   + i v(t)\sin{(\Delta\phi_{dy}(t)/2)} ,
\end{equation}
where
\begin{eqnarray}
 u(t) &=& \frac{1}{\sqrt{2}} \left(
     \cos{\frac{\theta_i-\theta(t)}{2}} - \sin{\frac{\theta_i-\theta(t)}{2}}
                             \right) , \\
 v(t) &=& \frac{1}{\sqrt{2}} \left(
     \cos{\frac{\theta_i+\theta(t)}{2}} - \sin{\frac{\theta_i+\theta(t)}{2}}
                             \right) .
\end{eqnarray}
\end{subequations}
$\theta$($\theta_i$) is the angle between the $z$-axis and
$\mathbf{B}$($\mathbf{B}_i$) in the Bloch sphere.

Fig.~4 displays the time evolution of the occupation number, $n_1$,
as a function of $\Delta\phi_{dy}(t)$, for an adiabatic change in
the AB phase of the form
$\phi(t) = \frac{\phi_i}{2} \left[ 1-\tanh{(\eta\,\Delta\phi_{dy}(t))} \right]$.
The initial phase $\phi=\phi_i$ at $\Delta\phi_{dy}(t)\ll -\eta^{-1}$
changes adiabatically
to $\phi=0$ at $\Delta\phi_{dy}(t) \gg \eta^{-1}$.  The parameter $\eta$
determines the rate of change.
As discussed above,
we have another condition $\phi_a(t)=\phi_b(t)=\phi(t)/2$
for a symmetric ring, which
takes Faraday's induction into account.
The time evolution of $n_1$ for
$\phi_i=2\pi$ with $\eta=0.04$ is plotted in Fig.~4 (solid line).
This is compared to the case of the
static AB phase, $\phi(t)=2\pi$ (dashed line).
The result shows that the adiabatic change of the AB phase from $2\pi$ to
zero indeed leads to out-of-phase oscillation of $n_1$.
This shift of the phase in $n_1$ results from the {\em Faraday phase},
$\phi^F_a=\phi^F_b=-\Delta\phi/2=\pi$, as shown in Eq.~(\ref{eq:Fab}).
Note that the AB effect does not play any role when $\phi$ changes by
$2\pi$.

The double-dot AB ring system is equivalent
to a single Cooper pair box (SCB) composed of two Josephson junctions with
an AB flux~\cite{nakamura99,makhlin01}, if the two QD states are replaced by
the two charge states in the SCB. The experiments described above
can be applied equally to a SCB.
It could be more easily realized with a SCB,
considering the recent progress made in controlling superconducting
qubits~\cite{makhlin01}.

At this stage, we are able to address the question raised
at the very beginning of this Letter: the inconsistency between the arbitrary
local phase in an AB loop and the possibility of its measurement with a QST.
Of course, a measurement of the local phase of a static AB ring is
meaningless since it is arbitrary.
However, for a complete QST, one should also change the localized AB flux.
What is measured during a QST (which inevitably involves a change of the flux)
is not an arbitrary local phase but the
{\em Faraday phase} induced by the change in the flux itself.

In conclusion, we have shown that the relative local phase at different
strengths of flux in an AB loop is not arbitrary but is instead
determined by Faraday's law
of induction. This is in strong contrast to the arbitrary local phase
factor of an AB loop which depends on the choice of gauge in
the vector potential. Faraday's induction provides
a geometric and nontopological
contributions to the local phase of a ring.
Flux-switching experiments for double-dot rings
have been proposed to verify the effect of the {\em Faraday phase}.
Measurement of the {\em Faraday phase} in our setup of a double-dot ring
is just one example of the very general nature of the problem.
It should be observable in various different types of
AB loops, which calls for further study.

This work was supported by
National Research Foundation of Korea under Grant No.~2009-0072595 and
No.~2009-0084606, and by LG Yeonam Foundation.

\begin{figure}
\includegraphics[width=3.3in]{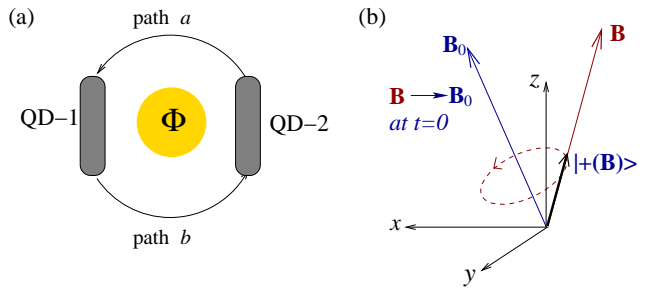}
\caption{(Color online) (a) Schematic of an Aharonov-Bohm ring composed of
two quantum dots.
(b) Illustration of the state evolution $|\psi_{\phi\rightarrow0}(t)\rangle$
in the Bloch-sphere representation after a sudden drop in the AB flux.}
\label{fig1}
\end{figure}

\begin{figure}
\includegraphics[width=1.1in]{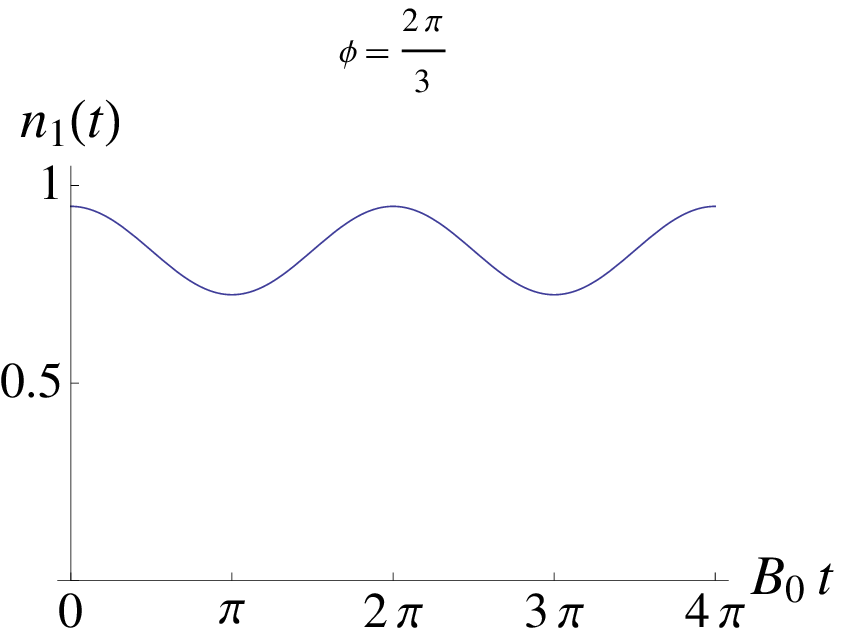}
\includegraphics[width=1.1in]{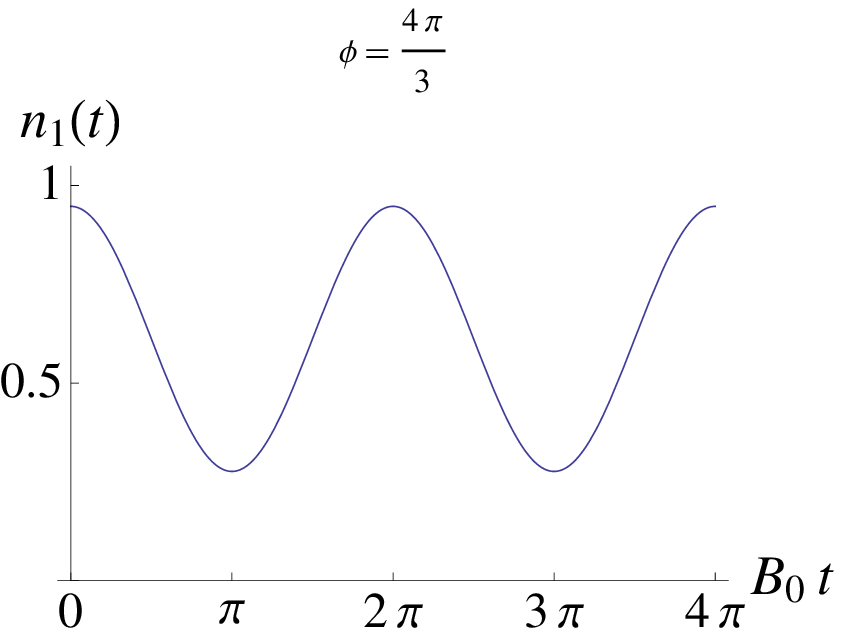}
\includegraphics[width=1.1in]{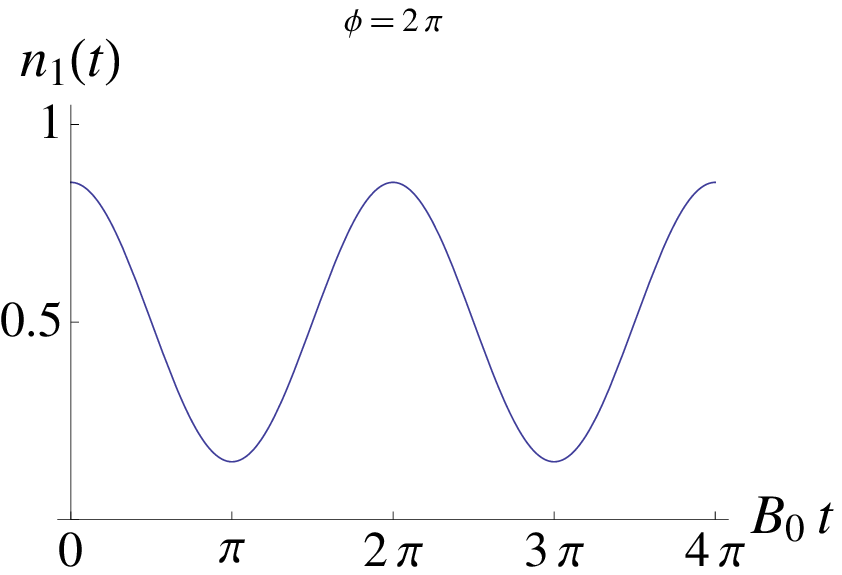}
\includegraphics[width=1.1in]{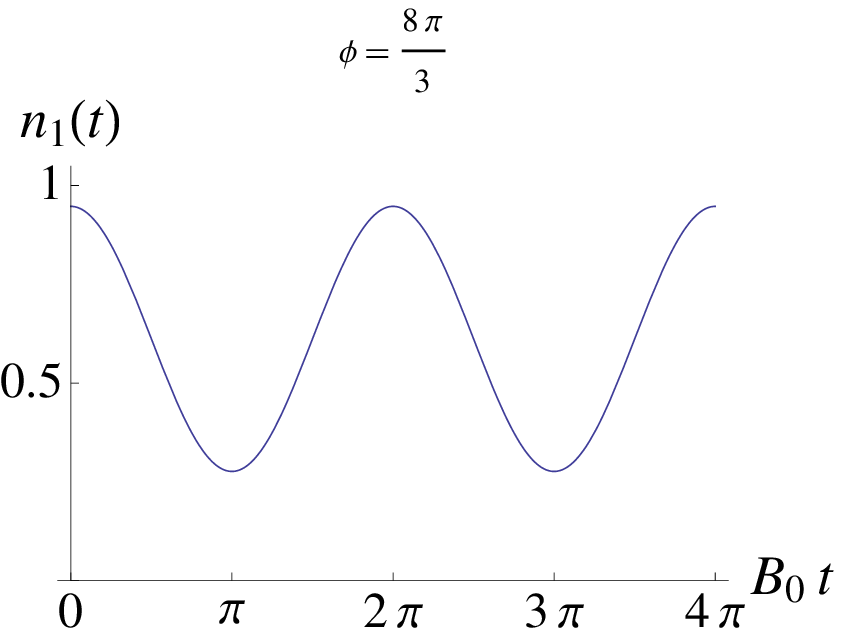}
\includegraphics[width=1.1in]{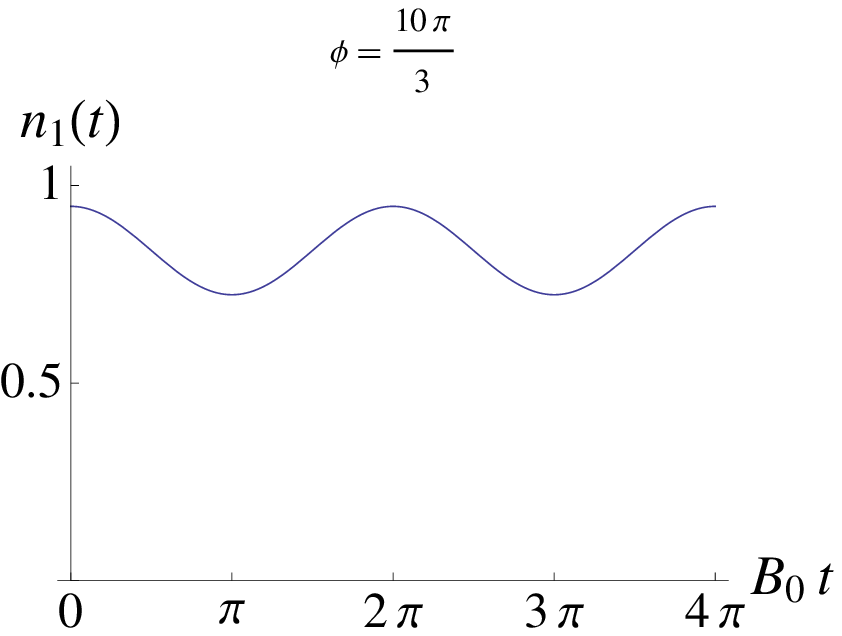}
\includegraphics[width=1.1in]{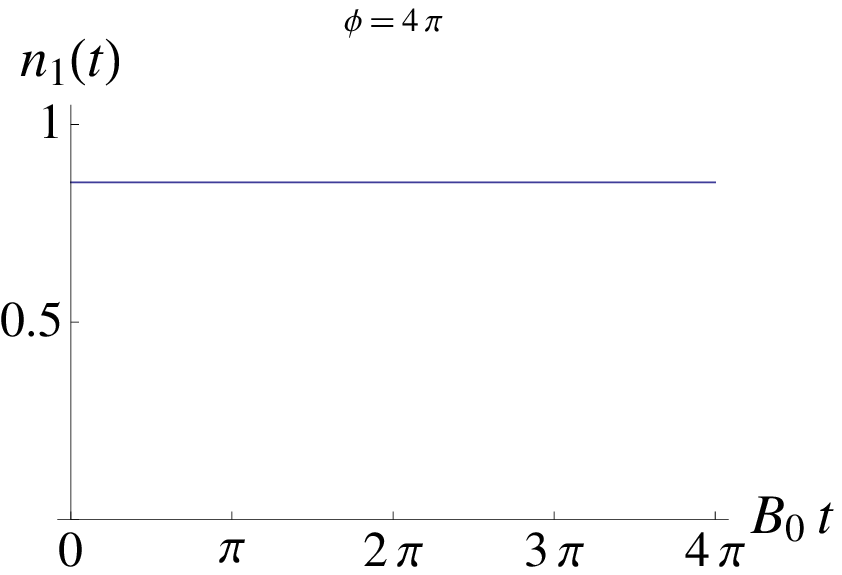}
\caption{(Color online) Time dependence of the electron number at QD-1
 ($n_1(t)$) for the {\em symmetric gauge},
 upon a sudden drop in the AB phase from $\phi$ to zero.
 Six different input values
 of $\phi$ are used, and $\Delta\varepsilon=4t$ in all graphs.}
 \label{fig2}
\end{figure}

\begin{figure}
\includegraphics[width=1.1in]{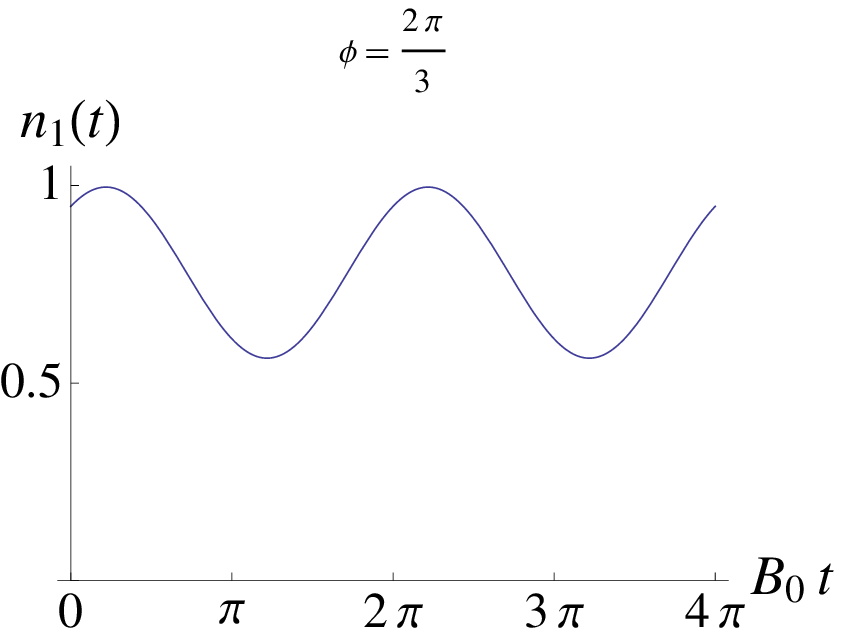}
\includegraphics[width=1.1in]{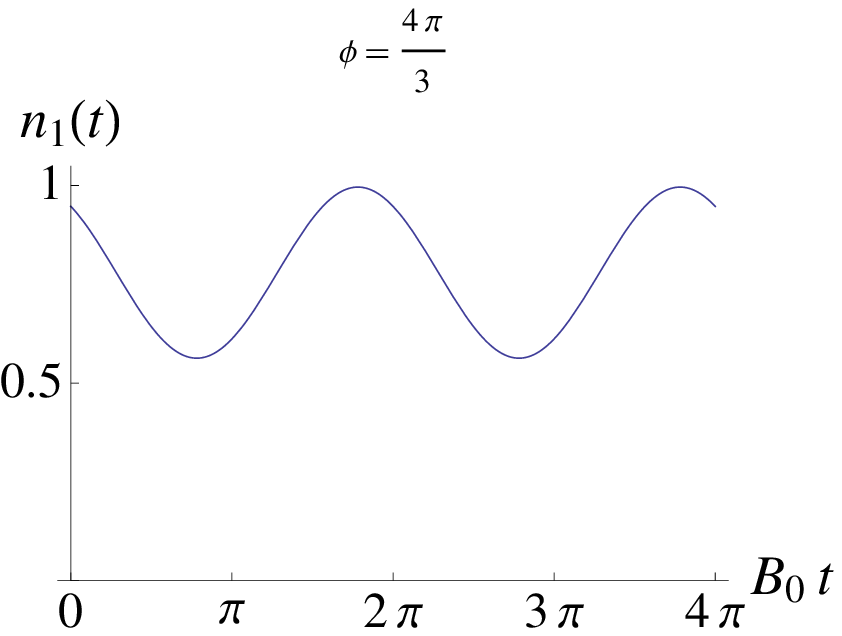}
\includegraphics[width=1.1in]{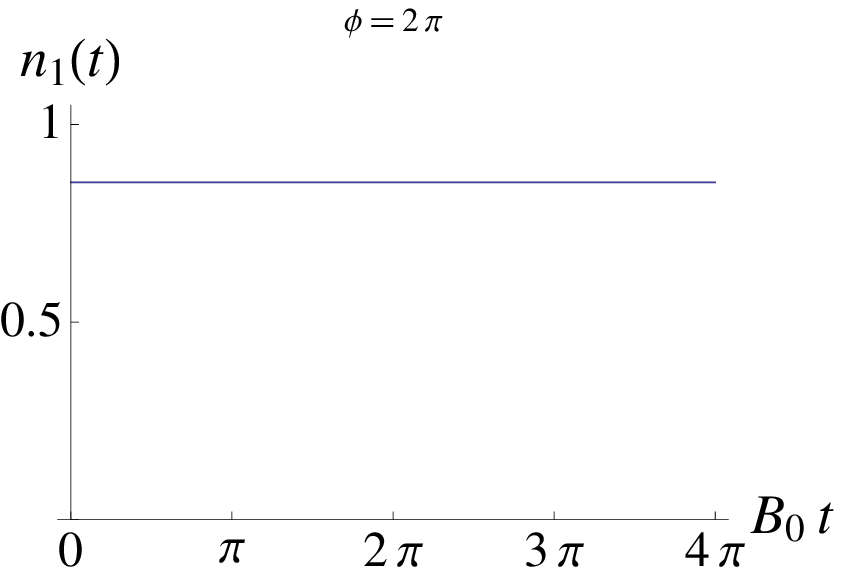}
\caption{(Color online) Time dependence of the electron number at QD-1
 ($n_1(t)$) for {\em ``$\phi_b=0$" gauge},
 upon a sudden drop in the AB phase from $\phi$ to zero. Three different input
 values of $\phi$ are used, and $\Delta\varepsilon=4t$ for all graphs.}
 \label{fig3}
\end{figure}
\begin{figure}
\includegraphics[width=3.0in]{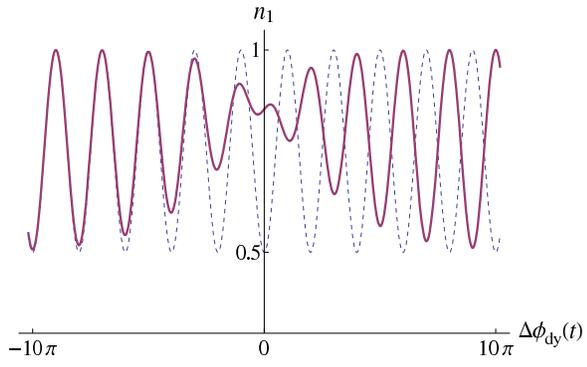}
\caption{(Color online) Time dependence of $n_1$ upon an adiabatic change of
the flux for a
nonstationary initial state of Eq.~(\ref{eq:psi0}) (solid line).
The parameters used here
are $\Delta\varepsilon=0\rightarrow 4t$, $\phi_i=2\pi$, and $\eta=0.04$.
For comparison, the time evolution of $n_1$ for a static AB phase ($\phi(t)=2\pi$)
is also plotted (dashed line).}
\label{fig4}
\end{figure}
\end{document}